\documentclass{pos}

\title{Measurement of the charge ratio of atmospheric muons with the CMS detector}

\ShortTitle{Measurement of the charge ratio of atmospheric muons with the CMS detector}

\author{
        Lars Sonnenschein$^a$ \\ 
        \scalebox{1.2}{\rm on behalf of the CMS collaboration} \\ \\
        $^a$III. Phys. Inst. A \\
        RWTH Aachen University\\
        E-mail: \email{Lars.Sonnenschein@cern.ch}
       }

\abstract{A measurement is presented of the flux ratio of positive and negative muons from 
        cosmic ray interactions in the atmosphere, using data collected by the CMS detector
        at ground level and in the underground experimental cavern. The excellent 
        performance of the CMS detector allowed detection of muons in the momentum range 
        from 5~GeV/c to 1~TeV/c. For muon momenta below 100~GeV/c the flux ratio is 
        measured to be a constant 
        $1.2766\pm 0.0032 (\mbox{stat.}) \pm 0.0032 (\mbox{syst.})$, 
        the most precise measurement to date. At higher momenta an increase in the charge 
        ratio is observed, in agreement with models of muon production in cosmic ray 
        showers and compatible with previous measurements by deep underground experiments. }

\FullConference{35th International Conference of High Energy Physics - ICHEP2010,\\
		July 22-28, 2010\\
		Paris France}

\begin{document}

\begin{figure}

\vspace*{-4.2ex}
\hspace*{6ex}
\includegraphics[width=12cm, bb= 100 400 500 580, clip]{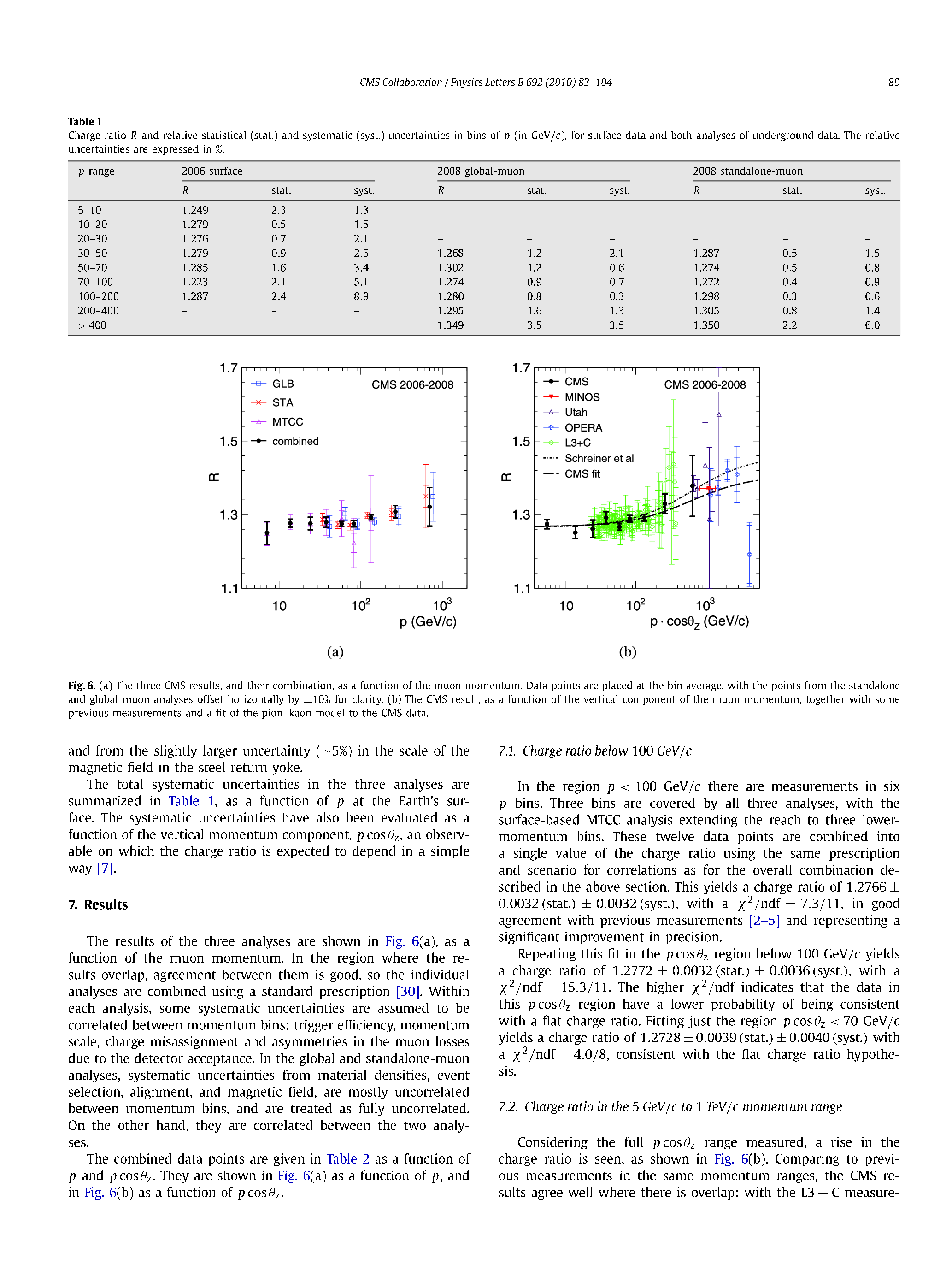}
\vspace*{-1.3ex}
\caption{\label{p_and_py}
  Charge ratio $R$ as a function of muon momentum (left)
  and its vertical component (right). Various measurements and fits are superposed.}
\end{figure}

\section{Dataset and measurement}
The flux ratio $R$ of positive to negative muons of cosmic ray 
interactions in the atmosphere is measured with the CMS detector~\cite{cmsdet}.
The Measurement is based on $3.3\cdot 10^5$ stand-alone muon track events taken 
with the muon drift tube chambers of the CMS detector at ground level in 2006 
as well as
$1.6\cdot 10^6$ stand-alone muon events and $2.45\cdot 10^5$ global muon events,
requiring a match to the central tracking system of the CMS detector, in the 
underground experimental cavern in 2008.
The underground measurements are corrected for muon energy loss in the material 
between earth's surface and the CMS detector. The muon momentum resolution is 
corrected by means of unfolding.

\section{Results}
In consistence with previous measurements (see Fig. \ref{p_and_py}) and with 
significant improvement in precision the charge ratio $R$ is found to be 
constant below muon momenta of $100$~GeV/c yielding
\begin{equation}
  R=1.2766 \pm 0.0032 (\mbox{stat.}) \pm 0.032 (\mbox{syst.}) .
\end{equation}
and below vertical muon momentum components of $70$~GeV/c yielding
\begin{equation}  
  R=1.2728 \pm 0.0039 (\mbox{stat.}) \pm 0.040 (\mbox{syst.}) .
\end{equation}
A rise of the charge ratio is seen (Fig. \ref{p_and_py}) above 
muon momenta of 100~GeV/c. The expected muon spectrum 
has been parameterised based on a pion kaon model \cite{gaisser}.
In comparison to the fit of Schreiner et al.~\cite{schreiner} a CMS fit
over the entire $p\cos\theta_z$ region yields the fractions of all pion 
and kaon decays into positive muons of
$f_{\pi}=0.533\pm0.005$ and $f_K=0.66\pm0.06$, respectively.


\begin{thebibliography}{99}


\bibitem{cmsdet} CMS collaboration,
 JINST {\bf 3}, S08004 (2008).

 \bibitem{c1} CMS collaboration, 
 Phys. Lett. {\bf B} 692, 83 (2010), CMS PAS MUO-10-001, CERN-PH-EP/2010-011, arXiv:1005.5332.

\bibitem{gaisser} T. Gaisser,
``Cosmic Rays and Particle Physics'', 
Cambridge University Press 1990.

\bibitem{schreiner} P. A. Schreiner et al.,
Astropart. Phys. {\bf 32, Issue 1}, 61 (2009).


\end{thebibliography}
\end{document}